\documentstyle[11pt]{article}
\def\v1{\vspace{1cm}}
\def\be{\begin{equation}}
\def\ee{\end{equation}}
\def\bc{\begin{center}}
\def\ec{\end{center}}
\def\ik{\partial}
\def\vh{\varphi}

\newcommand{\bea}{\begin{eqnarray}}
\newcommand{\eea}{\end{eqnarray}}

\begin{document}

\title
{\bf The Field Nature of Time in General Relativity}
\author{
M.~Pawlowski,\\
{\normalsize\it Soltan Institute for Nuclear Studies}, \\
{\normalsize\it Warsaw, Poland.}\\ [0.1cm]
V.~Pervushin, D.~Proskurin \\
{\normalsize\it Joint Institute for Nuclear Research},\\
{\normalsize\it 141980, Dubna, Russia.} }

\date{\empty}

\maketitle

\begin{abstract}
{{
The paper is devoted to the description of the
reparametrization - invariant dynamics of general relativity
obtained by resolving constraints and constructing equivalent
unconstrained systems. The constraint-shell action reveals the
"field nature" of the geometric time in general relativity. The
time measured by the watch of an observer coincides with one of
field variables, but not with the reparametrization-noninvariant
coordinate evolution parameter.

We give new solution of such
problem, as the derivation
of the path integral representation of the causal Green functions
in the Hamiltonian scheme of general relativity.
}}

\end{abstract}

{


\section{Introduction}

The problem of time and the reparametrization - invariant
Hamiltonian description of general relativity
has a long history~\cite{d1} -~\cite{ps1}.
There are two opposite approaches to solution of this problem in the
generalized Hamiltonian formulation~\cite{d1,d2,hrt,gt,kp,gkp}.

The first approach is
the reduction of the extended phase space by fixing the gauge that
breaks reparametrization invariance from the very beginning~\cite{ADM,ty}.

The second approach is the reparametrization-invariant reduction of
an action by the explicit resolving of the first
class constraints to get an equivalent unconstrained system,
so that one of the variables of the extended phase space
(with a negative contribution to the energy constraint) converts
into the {\it dynamic evolution parameter},
and its conjugate momentum becomes the nonzero Hamiltonian of
evolution~\cite{kuchar,grg,ps1,torre,kpp,pp,bp}.

An example of the application of such an invariant reduction of the action
is the Dirac formulation of QED~\cite{cj}
directly in terms of the gauge-invariant
(dressed) fields as the proof of the adequateness of the Coulomb gauge with
the invariant content of classical equations.
As it was shown by Faddeev~\cite{fad}, the invariant reduction of the action
is the way
to obtain the unconstrained Feynman integral for the foundation
of the intuitive Faddeev-Popov functional integral in the non-Abelian gauges
theories~\cite{fp2,fp1}.

The application of the invariant reduction
of extended actions in cosmology and general relativity~\cite{grg,ps1}
allows one to formulate the dynamics of relativistic systems
directly in terms of the invariant geometric time with the nonzero
Hamiltonian of evolution, instead of the non-invariant coordinate time with
the generalized zero Hamiltonian of evolution in the gauge-fixing method.
The formulation in terms of the geometric time is based on
the Levi-Civita canonical transformation~\cite{gkp,lc}
that converts the energy constraint into a new momentum, so that
the new dynamic evolution parameter coincides with the geometric time,
as one of the consequences of new equations of motion.

In the present paper, we apply the method of the invariant Hamiltonian
reduction (with resolving the first class constraints)
to express reparametrization-invariant dynamics of  relativistic systems
in terms of the geometric time and to construct the causal Green
functions in the form of the path integrals in the world space
of dynamic variables.

\section{Hamiltonian Dynamics of General Relativity}

\subsection{Action and geometry}

General relativity (GR) is
given by the singular Einstein-Hilbert action with the matter fields
\be
\label{gr}
W(g|\mu)= \int d^4x\sqrt{-g}[-\frac{\mu^2}{6} R(g)+{\cal L}_{matter}] ~~~
~~\left(~\mu^2=M^2_{Planck}\frac{3}{8\pi}~\right)
\ee
and by a measurable interval in the Riemannian geometry
\be \label{dse1}
  (ds)^2=g_{\alpha\beta}dx^\alpha dx^\beta~.
\ee
They are invariant with respect to general coordinate transformations
\be \label{x}
x_{\mu} \rightarrow  x_{\mu}'=x_{\mu}'(x_{0},x_{1},x_{2},x_{3}).
\ee

\subsection{Variables and Hamiltonian}

The generalized Hamiltonian approach to GR was formulated by Dirac
and Arnovit, Deser and Misner \cite{ADM} as a theory of system with
constraints in $3+1$ foliated space-time
\be
\label{dse}
  (ds)^2=g_{\mu\nu}dx^\mu dx^\nu= N^2 dt^2-{}^{(3)}g_{ij}\breve{dx}{}^i
  \breve{dx}{}^j\;~~~~~\;\;(\breve{dx}{}^i=dx^i+N^idt)
\ee
with the lapse function $N(t,\vec x)$, three shift vectors $N^i(t,\vec x)$,
and six space components ${}^{(3)}g_{ij}(t,\vec x)$ depending on the
coordinate time $t$ and the space coordinates $\vec x$.
The Dirac-ADM parametrization of metric~(\ref{dse}) characterizes
a family of hypersurfaces $t=\rm{const.}$ with the unit normal vector
$\nu^{\alpha}=(1/N,-N^k/N)$ to a hypersurface and with the second
(external) form
\be \label{ext}
\frac{1}{N}({}^{(3)}\dot g_{ij})-\Delta_i N_j -\Delta_j N_i)
\ee
that shows how this hypersurface is embedded into the four-dimensional
space-time.

Coordinate transformations conserving the family of
hypersurfaces $t=\rm{const.}$
\be \label{gt}
t \rightarrow \tilde t=\tilde t(t);~~~~~
x_{i} \rightarrow  \tilde x_{i}=\tilde x_{i}(t,x_{1},x_{2},x_{3})~,
\ee
\be \label{kine}
\tilde N = N \frac{dt}{d\tilde t};~~~~\tilde N^k=N^i
\frac{\partial \tilde x^k }{\partial x_i}\frac{dt}{d\tilde t} -
\frac{\partial \tilde x^k }{\partial x_i}
\frac{\partial x^i}{\partial \tilde t}
\ee
are called a kinemetric subgroup of the group of general coordinate
transformations (\ref{x})~\cite{vlad,grg,plb,ps1}.
The group of kinemetric transformations
is the group of diffeomorphisms of the generalized Hamiltonian dynamics.
It includes  reparametrizations of the nonobservable
time coordinate $\tilde t(t)$)~(\ref{gt})
that play the principal role in the procedure of the
reparametrization-invariant reduction discussed in the previous Sections.
The main assertion of the invariant reduction is the following:
the dynamic evolution parameter is not the coordinate but the variable
with a negative contribution to the energy constraint.
(Recall that this reduction is based on the explicit resolving of
the global energy constraint with respect to the conjugate momentum
of the dynamic evolution parameter to convert this momentum into
the Hamiltonian of evolution of the reduced system.)

A negative contribution to the energy constraint is given by
the space-metric-determinant logarithm. Therefore, following papers
\cite{kuchar,grg,torre,kpp,Y,yaf} we introduce an invariant evolution
parameter $\vh_0(t)$ as the zero Fourier harmonic component of this
logarithm (treated, in cosmology, as the cosmic scale factor).
This variable is distinguished in general relativity
by the Lichnerowicz conformal-type transformation of field
variables $f$ with the conformal weight $(n)$
\be \label{lich}
{}^{(n)}\bar f={}^{(n)}f \left(\frac{\vh_0(t)}{\mu}\right)^{-n}~,
\ee
where  $n=2,\;0,\;-3/2,\;-1$ for the tensor, vector,
spinor, and scalar fields, respectively, $\bar f$ is
so-called conformal-invariant variable used in GR
for the analysis of initial data~\cite{Y,L}.
In particular, for metric we get
\be \label{conf}
g_{\mu\nu}(t,\vec x)=
\left({\vh_0(t)\over \mu}\right)^2
\bar g_{\mu\nu}(t,\vec x)~\Rightarrow~
  (ds)^2=
\left({\vh_0(t)\over \mu}\right)^2
[\bar N^2 dt^2-{}^{(3)}\bar g_{ij}\breve{dx}{}^i \breve{dx}{}^j]~.
\ee
As the zero Fourier harmonic is extracted from the space metric determinant
logarithm, the space metric $\bar g_{ij}(t,\vec x)$ should be defined in
a class of nonzero harmonics
\be \label{gbar}
\int d^3x\log ||\bar g_{ij}(t,\vec x)||=0.
\ee
The transformational properties of the curvature $R(g)$ with respect to
the transformations~(\ref{conf}) lead to the action~(\ref{gr})
in the form~\cite{grg}
\be \label{Econf}
W(g|\mu)=W(\bar g|\vh_0) -\int\limits_{t_1 }^{t_2 }dt
\int\limits_{V_0}d^3x\vh_0
\frac{d}{dt}
(\frac{\dot \vh_0 \sqrt{\bar g}}{ \bar N}).
\ee
This form define the global lapse function $N_0$ as the average of the lapse
function $\bar N$ in the metric $\bar g$ over the kinemetric invariant space
volume
\be  \label{aven}
N_0(t)=\frac{V_0}{\int\limits_{V_0} d^3x \frac{\sqrt{\bar g(t, \vec x}}
{ \bar N(t, \vec x)}}~,~~~~~~~~~~~ \bar g=det({}^{(3)}\bar g)~,
~~~~~~~~~V_0=\int\limits_{V_0 }^{ }d^3x~,
\ee
where $V_0$ is a free parameter which in the perturbation theory
has the meaning of a finite volume of the free coordinate space.
The lapse function $\bar N(t, \vec x)$ can be factorized into the global
component $N_0(t)$ and the local one ${\cal N}(t, \vec x)$
\be \label{ncal1}
\bar N(t, \vec x) \bar g^{-1/2}:= N_0(t){\cal N}(t, \vec x):=N_q~,
\ee
where $\cal N$ fulfills normalization condition:
\be \label{ncal}
I[{\cal N}]:={1\over V_0}\int{d^3x\over {\cal N}}=1
\ee
that is imposed after the procedure of variation of action, to reproduce
equations of motion of the initial theory.
In the Dirac harmonical variables~\cite{d1} chosen as
\be \label{hv}
q^{ik}=\bar g \bar g^{ik},
\ee
the metric~(\ref{dse}) takes the form
\be \label{dse11}
(ds)^2=
\frac{\vh_0(t)^2}{\mu^2}
q^{1/2} \left(~N_q^2 dt^2 -q_{ij}\breve{dx}{}^i
  \breve{dx}{}^j \right),
~~~~~~~~ (q=det ( q^{ij} )) .
\ee
The Dirac-Bergmann version of action (\ref{Econf}) in terms of the introduced
above variables reads
~\cite{grg,plb}
\be \label{Egrc2}
W=
\int\limits_{t_1}^{t_2} dt \left\{
L+\frac{1}{2}\partial _t(P_0\vh_0)
\right\},
\ee
\be \label{Egrc3}
 L =
 \left[\int\limits_{V_0} d^3x
\left(\sum\limits_{F }^{ }P_F\dot F - N^i{\cal P}_i\right)\right] -
P_0 \dot\vh_0 - N_0\left[-\frac{P_0^2}{4V_0}+ I^{-1} H(\vh_0)\right]~,
\ee
where
\be \label{sum}
\sum\limits_{F }^{ }P_F\dot F = \sum\limits_{f }^{ }p_f\dot f-
\pi_{ij}\dot { q}^{ij};
\ee
\be \label{fpt1}
H(\vh_0)=\int d^3x{\cal N}{\cal H}(\vh_0)
\ee
is the total Hamiltonian of the local degrees of freedom,
\be \label{hh}
{\cal H}(\vh_0)=\frac{6}{\vh_0^2} q^{ij} q^{kl}[\pi_{ik}\pi_{jl}-
\pi_{ij}\pi_{kl}]
+\frac{\vh_0^2 q^{1/2}}{6}{}^{(3)}R(\bar g) +{\cal H}_f~,
\ee
and
\be \label{ph}
{\cal P}_i
=2[\nabla_k(q^{kl}\pi_{il})-\nabla_{i}(q^{kl}\pi_{kl})] +{\cal P}_{i f}~
\ee
are the densities of energy and momentum
and ${\cal H}_f,{\cal P}_f$ are contributions of the matter fields.
In the following, we call the set of the field variables $F$~(\ref{sum})
with the dynamic evolution parameter $\vh_0$ the field world space.
The local part of the momentum of the space metric determinant
\be \label{cs1}
\pi (t,x) :=  q^{ij}\pi_{ij}
\ee
is given in the class of functions with the non-zero Fourier harmonics,
so that
\be \label{cstr2}
\int d^3x  \pi (t,x)=0~.
\ee

The geometric foundation of introducing
the global variable ~(\ref{conf}) in GR was given in~\cite{yaf} as
the assertion about the nonzero value of the second form in the whole
space. This assertion (which contradicts the Dirac gauge $\pi=0$) follows
from the global energy constraint, as, in the lowest order of
the Dirac perturbation theory,
positive contributions of particle-like excitations to the
zero Fourier harmonic of the energy constraint can be compensated
only by the nonzero value of the second form.

\subsection{Local constraints and equations of motion}

Following Dirac~\cite{d1} we formulate generalized Hamiltonian
dynamics for the considered system (\ref{Egrc2}). It means the inclusion
of momenta for ${\cal N}$ and $N_i$ and appropriate terms with Lagrange
multipliers
\be \label{multipl}
W^D=
\int\limits_{t_1}^{t_2} dt \left\{L^D+\frac{1}{2}\partial _t(P_0\vh_0)
\right\},~~~
L^{D}=L+\int d^3x(P_{\cal N}\dot {\cal N} + P_{N^i}\dot N^i
- \lambda^0 P_{\cal N} - \lambda^i P_{N^i}).
\ee
We can define extended Dirac Hamiltonian as
\be \label{HED}
H^{D} = N_0\left[-\frac{P_0^2}{4V_0}+ I^{-1} H(\vh_0)
\right]+ \int d^3 x(\lambda^0 P_{\cal N} + \lambda^i P_{N^i}).
\ee
The equations obtained from variation of $W^D$ with respect to
Lagrange multipliers are called first class primary constraints
\be \label{primary}
P_{\cal N} =0,~~~~~~~~~~~~~~~~ P_{N^i}=0.
\ee
The condition of conservation of these constraints in time leads to
the first class secondary constraints
\be \label{secondary}
\left\{H^{D},P_{\cal N}\right\}={\cal H}-\frac{\int d^3x {\cal NH}}{V_0{\cal N}^2}=0,
~~~~~~~~~~
\left\{H^{D},P_{N^i}\right\}={\cal P}_i =0
\ee
For completeness of the system we have to include set of secondary
constraints. According Dirac we choose them in the form
\be \label{gauge1}
{\cal N}(t, \vec x) =1; ~~~~~~~~~~~ N^i(t, \vec x)  = 0;
\ee

\be \label{gauge2}
\pi(t, \vec x) =0; ~~~~~~~~~~~ \chi^j:=\partial_i(q^{-1/3}q^{ij})=0.
\ee
The equations of motion obtained for the considered system are
\be \label{eqED}
\frac{dF}{dT}=\frac{\partial H(\vh_0) }{\partial P_F },
~~
-\frac{dP_F}{dT}=\frac{\partial H(\vh_0) }{\partial F }~,
\ee
where $H(\vh_0)$ is given by the equation (\ref{fpt1}), and we
introduced the invariant geometric time $T$
\be \label{proptime}
N_0 dt:= dT~.
\ee

\subsection{Global constraints and equations of motion.}

The physical meaning of the geometric time $T$, the dynamic variable
$\vh_0$ and its momentum is given by
the explicit resolving of the zero-Fourier harmonic of the energy
constraint
\be \label{grconstr}
\frac{\delta W^E}{\delta N_0(t)}=
-\frac{P_0^2}{4V_0}+H(\vh_0) =0.
\ee
This constraint has two solutions for the global momentum $P_0$:
\be\label{glop}
(P_0)_{\pm}=\pm 2 \sqrt{V_0 H(\vh_0) }\equiv H^*_{\pm}.
\ee
The equation of motion for this global momentum $P_0$ in
gauge (\ref{gauge1}) takes the form
\be \label{nono}
\frac{\delta W^E}{\delta P_0}=0\,
\Rightarrow\,
\left(\frac{d\vh}{dT}\right)_{\pm}
=\frac{(P_0)_{\pm}}{2V}
=\pm\sqrt{\rho({\vh_0})};~~~\rho(\vh_0) =\frac{\int d^3x  {\cal H}}{ V_0}=
\frac{H(\vh_0) }{V_0}~.
\ee
The integral form of the last equation is
\be
\label{70}
T_{\pm}( {\vh_1},{\vh_0})=\pm
\int\limits_{\vh_1}^{\vh_0}d\vh{\rho}^{-1/2}(\vh)~,
\ee
where $\vh_1=\vh_0(t_1)$ is the initial data.
Equation obtained by varying the action with respect to $\vh_0$ follows
independently from the set of all other constraints and equations of motion.

In quantum theory of GR (like in quantum theories of a particle),
we get two Schr\"odinger equations
\be \label{sch1}
 i\frac{d}{d \vh_0}\Psi^{\pm}(F|\vh_0,\vh_1)=
H^*_{\pm}(\vh_0)\Psi^{\pm}(F|\vh_0,\vh_1)
\ee
with positive and negative eigenvalues of $P_0$
and normalizable wave functions with the spectral series
over quantum numbers $Q$
\be \label{psi+}
\Psi^+(F|\vh_0,\vh_1)=
\sum\limits_{Q }^{ }
A^+_Q <F|Q><Q|\vh_0,\vh_1>\theta (\vh_0-\vh_1)
\ee
\be \label{psi-}
\Psi^-(F|\vh_0,\vh_1)=
\sum\limits_{Q }^{ }
A_Q^-<F|Q>^*<Q|\vh_0,\vh_1>^*\theta (\vh_1-\vh_0)~,
\ee
where $<F|Q>$ is the eigenfunction of the reduced energy~(\ref{glop})
\be \label{eig}
H^*_{\pm}(\vh_0)<F|Q>=\pm E(Q,\vh_0)<F|Q>
\ee
\be \label{fq}
<Q|\vh_0,\vh_1>=\exp[-i\int\limits_{\vh_1 }^{\vh_0 }d\vh E(Q,\vh)],~~~~~~
 <Q|\vh_0,\vh_1>^*=\exp[i\int\limits_{\vh_1 }^{\vh_0 }d\vh E(Q,\vh)]~.
\ee
The coefficient $A^+_Q$, in "secondary" quantization, can be treated
as the operator of creation of a universe with positive energy;
and the coefficient $A^-_Q$, as the operator of
annihilation of a universe also with positive energy.
The "secondary" quantization means $[A_Q^-,A_{Q'}^+]=\delta_{Q,Q'}$.
The physical states of a quantum universe  are formed by the action of these
operators on the vacuum $<0|$, $|0>$ in the form of out-state
($|Q>=A_Q^+|0>$) with positive "frequencies" and in-state ($<Q|=<0|A_Q^-$)
with negative "frequencies".
This treatment means that positive frequencies propagate forward
(${\vh_0}>{\vh}_1$);
and negative frequencies, backward (${\vh}_1>{\vh}_0$), so that
the negative values of energy are excluded from the spectrum
to provide the stability of the quantum system in quantum theory of GR.
In other words, instead of  changing the sign of energy,
we change that of the dynamic evolution parameter, which leads to
the causal Green function
\be \label{causgr}
G_c(F_1,\vh_1|F_2,\vh_2)=G_+(F_1,\vh_1|F_2,\vh_2)\theta(\vh_2-\vh_1) +
G_-(F_1,\vh_1|F_2,\vh_2)\theta(\vh_1-\vh_2)
\ee
where $G_+(F_1,\vh_1|F_2,\vh_2)=G_-(F_2,\vh_2|F_1,\vh_1)$
is the "commutative" Green function
\be \label{FIpgr}
G_{+}(F_2,\vh_2|F_1,\vh_1)= <0|\Psi^-(F_2|\vh_2,\vh_1) \Psi^+(F_1|\vh_1,\vh_1)|0>
\ee
For this causal convention, the geometric time~(\ref{70})
is always positive in accordance with the equations of motion~(\ref{nono})
\be \label{atgr}
\left(\frac{d T}{d \vh_0}\right)_{\pm}=\pm \sqrt{\rho}~~
\Rightarrow~~T_{\pm}({\vh_1},{\vh_0})=
\pm\int\limits_{\vh_1}^{\vh_0}d\vh{\rho}^{-1/2}(\vh) \geq 0~.
\ee
Thus, the causal structure of the field world space immediately
leads to the arrow of the geometric time~(\ref{atgr})
and the beginning of evolution of a
universe with respect to the geometric time $T=0$.

The way to obtain conserved
integrals of motion in classical theory and quantum numbers $Q$
in quantum theory is the Levi-Civita-type canonical transformation
of the field world space $(F,\vh_0)$ to a geometric set of variables ($V,Q_0$)
with the condition that the geometric evolution parameter $Q_0$ coincides
with the geometric time $dT=dQ_0$.

Equations (\ref{nono}), (\ref{70}) in the homogeneous approximation of GR
are the basis of observational cosmology where the geometric time is the
conformal time connected with the world time $T_f$ of the Friedmann cosmology
by the relation
\be \label{fried}
dT_f=\frac{\vh_0(T)}{\mu}dT~,
\ee
and the dependence of scale factor (dynamic evolution parameter $\vh_0$)
on the geometric time $T$ is treated as the evolution of the universe.
In particular, equation (\ref{nono}) gives the relation between the
present-day value of
the dynamic evolution parameter $\vh_0(T_0)$ and cosmological observations,
i.e., the density of matter $\rho$ and the Hubble parameter
\be \label{mu}
{\cal H}_{hub}^e
=\frac{\mu\vh_0'}{\vh_0^2}=
\frac{\mu\sqrt{\rho}}{\vh_0^2}~~~~\Rightarrow ~~~~
\vh_0(T_0)=\left(\frac{\mu\sqrt{\rho}}{{\cal H}_{hub}}\right)^{1/2}:=
\mu\Omega_{0}^{1/4} ~
\ee
where $(~0.6< (\Omega_0^{1/4})_{exp}<1.2~)$.
The dynamic evolution parameter as the cosmic scale factor and a value of its
conjugate momentum (i.e., a value of the dynamic Hamiltonian) as the density
of matter (see equations (\ref{nono}), (\ref{70}))
are objects of measurement in observational astrophysics and cosmology
and numerous discussions about the Hubble parameter, dark matter, and hidden
mass.

The general theory of reparametrization-invariant reduction described
in the previous Sections can be applied also to GR. In accordance with this
theory, the reparametrization-invariant dynamics of GR is covered by
two unconstrained systems (dynamic and geometric) connected by the
Levi-Civita canonical transformation which solves the problems
of the initial data, conserved quantum numbers, and
direct correspondence of standard classical cosmology
with quantum gravity on the level of the generating functional
of the unitary and causal perturbation theory~\cite{ps1,pp}.

\section{Equivalent Unconstrained Systems}

Assume that we can solve the constraint equations and pass to the reduced
space of independent variables ($F^*,P_F^*$).
The explicit solution of the local and global constraints has two
analytic branches with positive and negative values for scale factor
momentum $P_0$ (\ref{glop}). Therefore, inserting solutions of all
constraints into the action we get two branches of the equivalent
Dynamic Unconstrained System (DUS)
\be \label{DUS}
W_{\pm}^{DUS}[F|\vh_0]
=\int\limits_{\vh_1}^{\vh_2} d{\vh_0}\left\{  \left[\int d^3 x
\sum_{F^*} P_F^* \frac{\partial F^*}{\partial \vh_0}\right] - H^*_{\pm}
+ \frac{1}{2} \partial_{\vh_0}(\vh_0 H^*_{\pm})\right\}~,
\ee
where $\vh_0$ plays the role of evolution parameter and $H^*_{\pm}$ defined
by equation (\ref{glop}) plays the role
of the evolution Hamiltonian, in the reduced phase
space of independent physical
variables $(F^*, P_F^*)$ with equations of motion
\be \label{EDUS}
\frac{dF^*}{d\vh_0}=\frac{\partial H^*_\pm}{\partial P_F^* },
~~~~~~~~
-\frac{dP_F^*}{d\vh_0}=\frac{\partial H^*_\pm}{\partial F^* }~.
\ee

The evolution of the field world space variables $(F^*,\vh_0)$ with respect
to the geometric time $T$ is not
contained in DUS (\ref{DUS}). This geometric time evolution is described by
supplementary equation~(\ref{nono}) for nonphysical momentum $P_0$ (\ref{glop})
that follows from the initial extended system.

To get an equivalent unconstrained system in terms of the geometric time
(we call it the Geometric Unconstrained System (GUS)), we need the
Levi-Civita canonical transformation (LC) \cite{gkp,lc} of the field world
phase space
\be \label{L-C}
(F^*,P^*_F|\vh_0,P_0)\Rightarrow (F_G^*,P_G^*|Q_0,\Pi_0)
\ee
which converts the energy constraint (\ref{grconstr}) into the new momentum
$\Pi_0$.

In terms of geometrical variables
the action takes the form
\be \label{L-CW}
W^G=\int\limits_{t_1}^{t_2}dt\left\{\left[\int d^3x\sum_{F_G^*}{P_G^*}
\dot F_G^*\right]- \Pi_0\dot Q_0+N_0\Pi_0+\frac{d}{dt}S^{LC}\right\}
\ee
where $S^{LC}$ is generating function of LC transformations.
Then the energy constraint and the supplementary equation for the new
 momentum take trivial form
\be \label{trivial}\Pi_0=0~;~~~~
\frac{\delta W}{\delta \Pi_0}=0 ~~~~\Rightarrow ~~~~\frac{dQ_0}{dt}=N_0 ~~~~ \Rightarrow ~~~~{dQ_0}={dT}.
\ee
Equations of motion are also trivial
\be \label{GUSE}
\frac{dP^*_G}{dT}=0,  ~~~~~~~~\frac{dF^*_G}{dT}=0,
\ee
and their solutions are given by the initial data
\be \label{initialG}
P^*_G=  {P^*_G}^0,~~~~~~~~F^*_G=  {F^*_G}^0.
\ee
Substituting solutions of (\ref{trivial}) and (\ref{GUSE}) into
the inverted Levi-Civita transformations
\be \label{L-C-reverse}
F^*=F^*(Q_0, \Pi_0|F^*_G, P^*_G), ~~~~~  \vh_0=\vh_0(Q_0, \Pi_0|F^*_G, P^*_G)
\ee
and similar for momenta, we get formal solutions of (\ref{EDUS})
and (\ref{70})
\be \label{sol-DUS}
F^*=F^*(T, 0|{F^*_G}^0, {P^*_G}^0), ~~~~~ P_F^*=P_F^*(T, 0|{F^*_G}^0, {P^*_G}^0),
~~~~~ \vh_0=\vh_0(T, 0|{F^*_G}^0, {P^*_G}^0).
\ee
We see that evolution of the dynamic variables with respect to the
geometric time (i.e., the evolution of a universe) is absent in DUS.
The evolution of the dynamic variables with respect to the geometric
time can be described
in the form of the  LC (inverted) canonical transformation
of GUS into DUS (\ref{L-C-reverse}), (\ref{sol-DUS}).

There is also the weak form of Levi-Civita-type transformations to GUS
$(F^*,P^*_F) ~~ \Rightarrow ~~ (\tilde F, \tilde P)$ without action-angle
variables and with a constraint
\be \label{weak-constraint}
\tilde\Pi_0-\tilde H(\tilde Q_0, \tilde F, \tilde P)=0.
\ee
We get the constraint-shell action
\be \label{weak-action}
\tilde W^{GUS}=\int dT \left\{\left[\int d^3x\sum_{\tilde F}\tilde P
\frac{d\tilde F}{dT}\right] - \tilde H(T,\tilde F, \tilde P)
\right\},
\ee
that allows us to choose the initial cosmological data with respect to the
geometric time.

Recall that the considered reduction of the action
reveals the difference of reparametrization-invariant theory
from the gauge-invariant theory: in gauge-invariant theory
the superfluous (longitudinal) variables are completely excluded
from the reduced system; whereas, in reparame-\\trization-invariant theory
the superfluous (longitudinal) variables leave  the sector of
the Dirac observables (i.e., the phase space ($F^*,P_F^*$))
but not the sector of measurable quantities: superfluous (longitudinal)
variables become the dynamic evolution parameter and dynamic Hamiltonian of
the reduced theory.

\section{Reparametrization-Invariant Path Integral}

Following Faddeev-Popov procedure we can write down the path
integral for local fields of our theory using constraints and gauge
conditions (\ref{primary}-\ref{gauge2}):
\be \label{idfp-local}
Z_{\rm local}(F_1,F_2|P_0,\vh_0,N_0)=\int\limits_{F_1}^{F_2}
D(F,P_f) \Delta_s \bar \Delta_t
\exp\left\{ i\bar W\right\},
\ee
where
\be \label{1prod}
D(F,P_f)=
\prod\limits_{t,x}
\left(\prod\limits_{i<k} \frac{dq^{ik}d\pi_{ik}}{2\pi}
\prod\limits_{f }^{ }\frac{df dp_f}{2\pi} \right)
\ee
are functional differentials for the metric fields ($\pi, q$) and the
matter fields ($p_f, f$),
\be \label{1fps}
\Delta_s=
\prod\limits_{t,x,i}
\delta({\cal P}_i)) \delta(\chi^j)det\{{\cal P}_i,\chi^j\} ,
\ee
\be \label{fpt}
\bar\Delta_t=
\prod\limits_{t,x}
\delta({\cal H}(\mu))\delta(\pi) det\{ {\cal H}(\vh_0)-\rho,\pi\},~~~~~~~~
\left(\rho=\frac{\int d^3x H(\vh_0)}{V_0}\right)
\ee
are the F-P determinants, and
\be \label{Egrc1}
\bar W=\int\limits_{t_1}^{t_2} dt \left\{\int\limits_{V_0} d^3x
\left(\sum\limits_{F }^{ }P_F\dot F \right)- P_0 \dot\vh_0
-N_0\left[-\frac{P_0^2}{4V_0}+  H(\vh_0)
\right]+\frac{1}{2}\partial _t(P_0\vh_0)
\right\}
\ee
is extended action of considered theory.

By analogy with a particle and a string considered in papers~\cite{pp,bp}
we define a commutative Green function as an integral over global fields
$(P_0,\vh_0)$ and the average over reparametrization group parameter $N_0$
\be \label{idfp}
G_+(F_1,\vh_1|F_2,\vh_2)=\int\limits_{\vh_1 }^{\vh_2 }
\prod\limits_{t }^{ }\left(\frac{d\vh_0 dP_0d\tilde N_0}{2\pi} \right)
Z_{\rm local}(F_1,F_2|P_0,\vh_0,N_0),
\ee
where
\be \label{average}
\tilde N={N}/{2\pi\delta(0)},~~~~~~ \delta(0)=\int dN_0.
\ee
The causal Green function in the world field space ($F,\vh_0$)
is defined as the sum
\be \label{cgfu}
G_c(F_1,\vh_1|F_2,\vh_2)=G_+(F_1,\vh_1|F_2,\vh_2)\theta(\vh_1-\vh_2)
+G_+(F_2,\vh_1|F_2,\vh_1)\theta(\vh_2-\vh_1).
\ee
This function will be considered as generating functional for
the unitary $S$-matrix elements ~\cite{bww}
\be \label{tr1}
S[1,2]=<{\mbox out}~(\vh_2)|
T_{\vh} \exp\left\{-i\int\limits_{\vh_1}^{\vh_2}d\vh (H^*_I)\right\}
|(\vh_1)~{\mbox in}>,
\ee
where $T_{\vh}$ is a symbol of ordering with respect to parameter $\vh_0$,
and $<{\rm out}~(\vh_2)|$,
$|(\vh)~ {\rm in }>$ are states of
quantum Universe in the lowest order of the Dirac perturbation theory
($ {\cal N}=1;~~N^k=0;~q^{ij}=\delta_{ij}+h^T_{ij} $),
 $H^*_I$ is the interaction Hamiltonian
\be \label{int}
H^*_I = H^*-H^*_0,~~~H^*=2\sqrt{V_0 H(\vh)}, ~~~~~H_0^*=2\sqrt{V_0 H_0(\vh)}~,
\ee
$H_0$ is a sum of the Hamiltonians of "free" fields
(gravitons, photons,  massive vectors, and spinors) where  all masses
(including the Planck mass) are replaced by the
dynamic evolution parameter $\vh_0$~\cite{ps1}. For example for gravitons
the "free" Hamiltonian takes the form:
\be \label{hom}
H_0(\vh_0)=\int d^3x\left(\frac{6(\pi^T_{(h)})^2}{\vh_0^2}+\frac{\vh_0^2}{24}
(\ik_ih^T)^2\right);~~(h^T_{ii}=0;~~\ik_jh^T_{ji}=0).
\ee

\subsection{QFT limit of  Quantum Gravity}

The simplest way to determine the QFT limit of  Quantum Gravity
and to find the region of
validity of the FP-integral is to use the quantum field
version of the reparametrization-invariant
integral(\ref{idfp}) in the form of
S-matrix elements ~\cite{bww} (see (\ref{tr1}), (\ref{int})).
We consider the infinite volume limit of
the S-matrix element (\ref{int}) in terms
of the geometric time $T$ for the present-day
stage  $ T=T_0, \vh(T_0)=\mu$, and
$T(\vh_1)=T_0-\Delta T, T(\vh_2)=T_0+\Delta T=T_{out}$.
One can express this matrix element in terms of the time measured by
an observer of an out-state with a tremendous number of particles
in a universe using equation $d\vh=dT_{out}\sqrt{\rho_{out}}$
and approximation of a tremendous energy $(10^{79} GeV)$ in
comparison  with possible real and virtual deviations of the free
Hamiltonian in the laboratory  processes:
\be \label{lab1}
 \bar H_0 = E_{out} + \delta H_0 ,~~~
<{\mbox out}| \delta H_0|{\mbox in}> << E_{out}.
\ee
to neglect "back-reaction".

In the infinite volume limit, we get from (\ref{int})
\be \label{limit} d \vh_0 [ H^*_I ]=2 d \vh_0 \left(\sqrt{V_0(H_0+H_I)}-
\sqrt{V_0 H_0}\right)=
dT_{out}[\hat F \bar H_I +  O(1/E_{out})]
\ee
where $ H_I$ is the interaction Hamiltonian in GR,
and
\be \label{form}
\hat F = \sqrt{\frac{E_{out}}{ H_0}}
= \sqrt{\frac{E_{out}}{E_{out} + \delta  H_0}}
\ee
is a multiplier which plays the role of a form factor for
physical processes  observed in  the "laboratory"  conditions
when the cosmic energy $E_{out}$ is much greater than the
deviation of the free energy
\be \label{lab}
\delta  H_0 =  H_0-E_{out};~~
\ee
due to creation and annihilation of
real and virtual particles in the laboratory experiments.

The measurable time of the laboratory experiments $T_2-T_1$ is much
smaller than the age of the universe $T_0$, but it is much greater
than the reverse "laboratory"  energy  $ \delta$, so that the
limit
$$
\int\limits_{T(\vh_1) }^{T(\vh_2) }dT_{\rm out}
\Rightarrow \int\limits_{-\infty }^{+\infty }dT_{\rm out}
$$
is valid.
If we neglect  the form factor ~(\ref{form}) that removes a set of
ultraviolet divergences, we get the standard S-matrix element~\cite{fp2}
that corresponds to the standard FP functional integral
with the geometric (conformal) time $T$ (instead of the coordinate time $t$)
and with conformal-invariant fields
$t \rightarrow T_{\rm out}$:
\be \label{tr4}
S[-\infty|+\infty]=<{\mbox out}|
T \exp\left\{-i\int\limits_{-\infty}^{+\infty} dT_{\rm out}\hat F
H_I(\mu)\right\}
|{\mbox in}> ~~~~~~~~ \left(\hat F=1\right).
\ee
Thus, the standard FP-integral and the unitary S-matrix for conventional
quantum field theory (QFT)
appears as the nonrelativistic
approximation of tremendous mass of a universe and its very large life-time.
Now, it is evident that QFT
are not valid for the description of the early universe given
in the finite spatial volume and the finite positive interval
of geometrical time $(0 \leq T \leq T_0)$ where $T_0$ is the "present-day
value" for the early universe that only begins to create matter.

On the other hand, we revealed that standard QFT (that appears as the limit
of quantum theory of the Einstein general relativity)
speaks on the language of the  conformal fields and coordinates.
If we shall consider the standard QFT as the limit case of quantum gravity,
we should recognize that, in QFT, we measure the conformal quantities, as
QFT is expressed in terms of the conformal-invariant Lichnerowicz variables
and coordinates including the conformal time ($T_{out}$) as the time
of evolution of these variables.

The conformal invariance of the variables can testify to the
conformal invariance of the initial theory of gravity of the type
of a dilaton version of GR given in a space with the geometry of
similarity~\cite{grg,plb}.

\section{ Conclusions}

All relativistic theories, including general
relativity considered in the present paper, are given
in their world spaces of dynamic variables by their singular actions
(as integrals over the coordinate space) and by the geometric interval.

The peculiarity of general relativity is
the invariance of its action and the geometric interval with respect to
reparametrization of the coordinate space, i.e., the general coordinate
transformations.

The main peculiarity of general relativity (which we tried to reveal
in the paper) is the following: the reparametrization symmetry means that
the measurable geometric time is a time-like variable in the
field world space (obtained by the Levi-Civita transformation to
the action-angle-type variables) rather than the coordinate.

The dynamic origin of the "time" was reliably covered
by the gauge condition that the lapse-function is equal to unity.

This non-invariant gauge-fixing method of describing the Hamiltonian
dynamics of general relativity  was a real obstacle for understanding
this dynamics.
This non-invariant method confuses reparametrization-invariant
(or measurable) quantities and non-invariant  (nonobservable) ones
and hides the necessity of constraining by the Levi-Civita transformation
that converts  ambiguous and attractive "mathematical games"
with non-invariant quantities into a harmonious theory of invariant
dynamics in the world space which includes an unambiguous
description of quantum gravity with its relation to the standard
cosmology of a classical universe.

The generating functionals for
causal Green functions of the unitary perturbation theory in
the form of  path integrals are constructed by averaging over
a space of the reparametrization group, instead of the gauge-fixing.

{\bf Acknowledgments}

\medskip

We are happy to acknowledge interesting and critical discussions with
V.G. Kadyshevsky, H. Kleinert, E. A. Kuraev,
V. I. Smirichinski, and I. V. Tyitin.


\vspace{1cm}

\begin{thebibliography}{}
\bibitem{d1}
Dirac P A M 1958 {\it Proc.Roy.Soc.} {\bf A 246} 333;
 1959 {\it Phys. Rev.} {\bf 114} 924
\bibitem{ADM}
Arnovitt R, Deser S and Misner C W 1959 {\it Phys. Rev.} {\bf 116} 1322;
 1960 {\it Phys. Rev.} {\bf 117} 1595
\bibitem{kuchar}
Isham C and Kuchar K 1985 {\it Ann. Phys. (NY)} {\bf 164} 288, 316
\bibitem{grg}
Gyngazov L N, et. al. 1998, {\it Gen. Rel. and Grav.} {\bf 30} 1749
\bibitem{plb}
Pawlowski M,  Papoyan V V, Pervushin V N,
 Smirichinski V I 1998 {\it Phys. Lett. } {\bf 444B } 293
\bibitem{ty}
Fulop G, Gitman D M and Tyutin I V  1999 {\it Int.J.Theor.Phys.} {\bf 38} 1941
\bibitem{ps1}
Pervushin V N and Smirichinski V I 1999
{\it J.Phys.A: Math.Gen.}{\bf 32} 6191
\bibitem{d2}
Dirac P A M 1964
{\it Lectures on Quantum Mechanics} (Belfer Graduate School of Science,
Yeshive University Press, New York)
\bibitem{hrt}
Hanson A J, Regge T and Teitelboim C 1976 {\it Constrained Hamiltonian
Systems} Accademia Nazionale dei Lincei (Rome)
\bibitem{gt}
Gitman D M and Tyutin I V  1990 {\it Quantization
of fields  with Constraints}
(Springer-Verlag, Berlin)
\bibitem{kp}
Konopleva N P and Popov V N 1972 {\it Gauge fields} (Moscow, Atomizdat)
(in Russian)
\bibitem{gkp}
Gogilidze S A, Khvedelidze A M and Pervushin V N 1999
{\it Phys.Particles and Nuclei} {\bf 30} 66
\bibitem{torre}
Torre C G 1991 {\it Class. Quantum Grav.} {\bf 8} 1895
\bibitem{kpp}
Khvedelidze A, Palii Yu, Papoyan V and Pervushin V 1997
{\it Phys. Lett.} {\bf 402 B} 263
\bibitem{pp}
Pawlowski M and Pervushin V N 2001
{\it Int.J.Mod.Phys.A} to be published;~hep-th/0006116
\bibitem{bp}
Barbashov B M and Pervushin V N 2001 {\it J.Phys.A} to be published;~
 hep-th/0005140
\bibitem{cj} Dirac P A M 1955  {\it Can. J. Phys.} {\bf 33} 650
\bibitem{fad} Faddeev L 1969 {\it Teor. Mat. Fiz.} {\bf 1} 3 (in Russian)
\bibitem{fp2}
Faddeev L D and Popov V N 1973 {\it Us.Fiz.Nauk} {\bf 111} 427
\bibitem{fp1} Faddeev L and  Popov V 1967
{\it Phys. Lett.} {\bf 25 B} 29;
DeWitt B S 1967  {\it Phys. Rev.} {\bf 160} 1113
\bibitem{lc}
Levi-Civita T 1906 {\it Prace Mat.-Fiz.} {\bf 17} 1;
Shanmugadhasan S 1973  {\it J. Math. Phys} {\bf  14\/} 677
\bibitem{vlad}
Zel'manov A L 1976 {\it Doklady AN SSSR} {\bf 227} 78 (in Russian);\\
Vladimirov Yu S 1982 {\it Frame of references in
theory of gravitation} (Moscow, Eneroizdat, in Russian)
\bibitem{Y}
York J W (Jr.) 1971 {\it Phys. Rev. Lett.} {\bf 26} 1658;~
Kuchar K 1972 {\it J. Math. Phys.} {\bf 13} 768
\bibitem{yaf}
Pervushin V N and Smirichinski V I 1998 {\it Physics of Atomic Nuclei}
{\bf 61} 142
\bibitem{L}
Lichnerowicz A 1944 {\it Journ. Math. Pures and Appl.}
\bibitem{bww}
Schweber S 1961 {\it An Introduction to Relativistic Quantum Field Theory}
(Row, Peterson and Co $\bullet$ Evanston, Ill., Elmsford, N.Y)
\end{thebibliography}
\end{document}